\documentclass[aps,twocolumn,prb,superscriptaddress]{revtex4}

\usepackage{graphicx}
\usepackage{amssymb}

\newcommand{\R}{\mathbf{r}}
\newcommand{\HEG}{{\rm HEG}}
\newcommand{\tk}{\tau^{\rm KS}}
\newcommand{\tw}{\tau^{\rm W}}
\newcommand{\RP}{{\mathbf{r}'}}

\begin{document}

\title{Kinetic and Exchange Energy Densities near the~Nucleus}
\author{Lucian A. Constantin}
\affiliation{Center for Biomolecular Nanotechnologies @UNILE, Istituto Italiano di Tecnologia, Via Barsanti, 
I-73010 Arnesano, Italy}
\author{Eduardo Fabiano}
\affiliation{Istituto Nanoscienze-CNR, Euromediterranean Center for Nanomaterial Modelling and Technology (EC
MT), via Arnesano 73100, Lecce}
\affiliation{Center for Biomolecular Nanotechnologies @UNILE, Istituto Italiano di Tecnologia, Via Barsanti, 
I-73010 Arnesano, Italy}
\author{Fabio Della Sala}
\affiliation{Istituto Nanoscienze-CNR, Euromediterranean Center for Nanomaterial Modelling and Technology (EC
MT), via Arnesano 73100, Lecce}
\affiliation{Center for Biomolecular Nanotechnologies @UNILE, Istituto Italiano di Tecnologia, Via Barsanti, 
I-73010 Arnesano, Italy}

\date{\today}

\begin{abstract}
We investigate the behavior of the kinetic and the exchange energy densities near the nuclear cusp of atomic systems.
Considering hydrogenic orbitals, we derive analytical expressions near the nucleus, for single shells, as well as in the semiclassical limit of large non-relativistic neutral atoms. 
We show that a model based on the helium iso-electronic series is very accurate, as also confirmed
 by numerical calculations
on real atoms up to two thousands electrons.
Based on this model, we propose non-local density-dependent ingredients that are 
suitable for the description of the kinetic and exchange energy densities in the region close to the nucleus.
These non-local ingredients are invariant under the uniform scaling of the density, 
and they can be used in the construction of non-local exchange-correlation and
kinetic functionals.
\end{abstract}

\maketitle

\section{Introduction}
Kohn--Sham (KS) density functional theory (DFT) \cite{KS,dreizbook,parrbook}
can be considered the most used method in electronic calculations of quantum
chemistry and condensed matter physics. Its practical implementation
is based on approximations of the exchange-correlation (XC) energy ($E_{xc}$),
which is a~subject of intense research \cite{scusrev,becke14,peveratirev}.
Moreover, subsystem DFT \cite{wesochemrev,jacobrev,pavanello15} 
and orbital-free DFT 
\cite{wangcarterof,wesobook,xia12,karasiev12} need the use 
of kinetic energy (KE) functional approximations. 

Concerning XC functionals, the simplest ones beyond the local density approximation (LDA)
are~those based on the generalized gradient approximation (GGA) 
\cite{pbe,b88,revpbe,rpbe,wc,htbs,apbe,pbemol,peverati12,am05,pbesol,vmt,pbeint,zpbeint,q2D,sogga11,sg4},
which are constructed using the electron density ($\rho$) and its gradient ($\nabla\rho$).
Meta-generalized gradient approximations (meta-GGAs)~\cite{tpss,revtpss,mggams1,sunpnas,m06l,m11l,bloc,b97mv2015,vt84,mn12l,beefmeta14,scan}
are the most sophisticated semilocal functionals, incorporating important
exact conditions and having an improved overall accuracy with respect to the GGA functionals.
The~meta-GGA functionals use as an additional ingredient to the GGA ones the
Kohn--Sham positive KE density: 
\begin{equation}
\tk(\R)=\frac{f}{2}\sum_{i=1}^N|\nabla\phi_{i}(\R)|^2 \; 
\label{defk}
\end{equation}
(the total KE being $T=\int \tk d^3\R$). In Equation (\ref{defk}), $f=2$ for closed-shell systems 
(as the ones considered in this work), and the summation is over all occupied orbitals; 
atomic units, \emph{i.e.}, $e^2=\hbar=m_e=1$, are~used throughout.
The quantity $\tk$ enters in
the expansion of the angle-averaged exact exchange hole \cite{Beckeh,scusrev}, being thus a natural and important 
tool in the construction of XC approximations.

An important requirement for an accurate XC functional is a proper model for the exchange~energy (XE) density \cite{b88,arm13}.
The XE density is usually defined in terms of the exchange enhancement factor:
\begin{equation}
F_x=\frac{e_x}{e_x^\HEG} \; 
\end{equation}
where $e_x^\HEG=-(3/(4\pi))(3\pi^2\rho)^{1/3}\rho$ is the XE density of the homogeneous electron gas (HEG). 
We recall that the XE density is not uniquely defined (being up to 
a gauge transformation), but its underlining hole must be realistic and close to the exact one, which 
is an observable \cite{blochole,perdewPRB96,ernzerhofJCP98,vydrov2006importance}. 
In this work, we will use as reference the definition of the conventional exact-exchange density.
At the GGA level, the total exchange energy is usually expressed as:
\begin{equation}
E_x^{{\rm GGA}}  = \int {\rm d}^3\R\, e_x^\HEG F_x (s) \; 
\end{equation}
\emph{i.e.}, with the exchange enhancement factor being a function of the reduced gradient
$s=|\nabla \rho|/[2 k_F \rho]$.
Here, $k_F=(3\pi^2\rho)^{1/3}$ is the local Fermi wavevector.
At the meta-GGA level, we have:
\begin{equation} 
E_x^{{\rm meta-GGA}}=\int {\rm d}^3\R\, e_x^\HEG F_x (s,q,\tk) \; 
\end{equation}
\emph{i.e.}, $F_x$ is also a function of reduced Laplacian
$q=\nabla ^2\rho/[4 k_F^2 \rho ]$ and/or of $\tk$.

At the nucleus of the helium isoelectronic series, it has been shown \cite{tao01} that 
$F_x=\frac{1}{3}(\frac{4\pi^2}{3})^{1/3}\approx 0.787$.
However, in popular semilocal exchange functionals (GGAs and meta-GGAs), $F_x\ge 1$: thus,~these functionals 
cannot be realistic near the nucleus, where there is an important de-enhancement. 
The~nuclear region can be identified using the usual semilocal ingredients. For example, the reduced gradient
$s$ behaves at the nucleus of the helium isoelectronic series 
as $s=1/(6\pi)^{1/3}\approx 0.376$, while the reduced Laplacian
$q$ diverges to $-\infty$. 
This issue has been considered by Tao 
\cite{tao01}, who constructed an exchange functional with the correct XE density at the nucleus,
using the single inhomogeneity parameter proposed by Becke \cite{Beckeh} $Q_B=1-\frac{\tau}{\tau^\HEG}+\frac{5s^2}{3}+\frac{10}{3}q$ with 
$\tau^\HEG=(3/10)(3\pi^2)^{2/3}\rho^{5/3}$.
Further~improvements on the development of Laplacian-dependent exchange functionals have been found by Cancio \emph{et al.} 
\cite{cancioIJQC12}.

Similar shortcomings as for the XE density near the nucleus affect also many KE functionals 
at the GGA level \cite{lc94,tw02,llp91,thak92,apbek} or
at the Laplacian level \cite{fde_lap}; for a recent review of semilocal functionals, \linebreak see \cite{weso_chap_funct}.
The KE density is usually defined in terms of the KE enhancement factor:
\begin{equation}
F_s=\frac{\tau}{\tau^\HEG}\; 
\end{equation} 
so that the total kinetic energy is: 
\begin{equation}
T_s=\int {\rm d}^3\R \, \tau^\HEG F_s(s,q) \; 
\end{equation} 

The von Weizs\"{a}cker (VW) kinetic energy density:
\begin{equation}
\tw[\rho]=\frac{|\nabla \rho|^2}{8\rho} \; 
\end{equation} 
(\emph{i.e.}, $F_s^W=(5/3)s^2$) is expected to be very accurate at the nucleus \cite{dreizbook,alva08,kara09,lastra08}. 
At the nucleus of the helium isoelectronic series, $F_s\approx F_s^W\approx 0.2353$, \emph{i.e.}, there is 
an even more pronounced de-enhancement than in the exchange case. On the other hand, most of the GGA KE functionals
have $F_s\ge 1$, while few semilocal KE functionals recover the VW at the nucleus \cite{MGGA,kara09,kara13}.

However, we recently pointed out that the VW functional does not have the correct behavior at the nucleus \cite{alpha}. 
It was proven that the KE density at the nuclear cusp behaves as \cite{alpha}: 
\begin{equation}
\tk=\tw[\rho_{s}]+ 3\tw[\rho_{p}] \; 
\label{eq1}
\end{equation}
where 
$\rho_{s}$ and $\rho_{p}$ are the densities of $s$-type and $p$-type shells, respectively. Thus, also
$p$-shells contribute to the KE density at the nuclear cusp \cite{alpha,qian07}. 
The second term on the right-hand-side of Equation (\ref{eq1}) has been evaluated for real atoms, and its contribution 
in the semiclassical limit of a neutral atom with an infinite number of electrons reaches $12\%$ of the total KE density
\cite{alpha}. 

Understanding the physical phenomena at the nucleus can thus boost the development of more accurate XC and KE approximations. 
We recall that the nucleus region contains an important 
part of the total kinetic and exchange energies, and thus, small modifications of the KE and XC enhancement 
factors can bring significant variations to total energies. 

In this paper, we will consider different aspects of density functionals in the nuclear regions:\linebreak 
(i) we will describe the difference between the exact and VW kinetic energy densities, in a region near the nucleus, extending
the derivation of \cite{alpha}, where only the nuclear cusp was considered;
(ii)~we will present an approach based on the helium iso-electronic 
series, which correctly describes the nuclear region, for small atoms up to the semiclassical limit of large non-relativistic neutral atoms;
(iii) we will propose novel non-local density ingredients for the 
conventional exchange and kinetic energy densities at the nuclear region.

The paper is organized as follows:
In Section \ref{sec:ke}, we present a detailed analysis of the kinetic and exchange energies at the nucleus of spherical systems. 
We investigate the hydrogenic shells, \linebreak the ten-electron hydrogenic model and the asymptotic neutral atom with an infinite number
of electrons, presenting the small-$r$ expansions of various quantities of interest. 
We also demonstrate that the 1$s$-shell model (1SM) approach is 
remarkably accurate for both kinetic and exchange energies, in the case of real atoms. 
The simple 1SM cannot be described by any semilocal ingredient, having a~significant amount of non-locality.
Consequently, in Section \ref{sec:nl}, we propose non-local ingredients for exchange and kinetic energies, near the nucleus. 
These approximations are invariant under the uniform scaling of the density, and they can be used in the construction of non-local 
functionals. 
Finally, in Section \ref{sec:cl}, we summarize the results.

\section{Kinetic and Exchange Energy Densities at the Nuclear Cusp in Spherical Systems} \label{sec:ke}
For a system in a central potential, the KS orbitals
can be written as $\phi_{nlm}(\R)=R_{nl}(r)Y_{lm}(\theta,\phi)$, where 
$R_{nl}(r)$ are the normalized radial functions, 
$Y_{lm}(\theta,\phi)$ are spherical harmonics, 
$n$ is the principal quantum number, $l$ is the angular momentum and $m$ is
the azimuthal quantum number. 
The density of the shell $nl$ is:
\begin{eqnarray}
\nonumber
\rho_{nl}(r) & = & f\sum_m \left|\phi_{nlm}(\R)\right|^2 = f\left|R_{nl}(r)\right|^2\sum_m \left|Y_{lm}(\theta,\phi)\right|^2 \\
\label{e21}
& = & f\left|R_{nl}(r)\right|^2\frac{2l+1}{4\pi} \; 
\end{eqnarray}
where in the last equality, we used Uns\"old's theorem for spherical harmonics.
The Kohn--Sham KE density of the shell $nl$ satisfies the relation \cite{alpha,nagy89,santamaria90}:
\begin{equation}
\tk_{nl}=\tau^W[\rho_{nl}]+\frac{l(l+1)}{2}\frac{\rho_{nl}}{r^2} \; 
\label{eq2}
\end{equation}
which is valid at any radial distance $r$ from the nucleus.

Note that the above relations are valid for systems in which a single shell is occupied.
For real systems, with many occupied shells, the total density and the total KE density are obtained summing over all shells, \emph{i.e.}:
\begin{eqnarray}
\rho &=&  \sum_{nl} \rho_{nl} \label{eq:rs} \; \\
\tau &=&  \sum_{nl} \tau_{nl} \label{eq:ts} \; 
\end{eqnarray}

Note that, instead, in general, $\tau^W \neq \sum_{nl} \tau^W_{nl}$ due to the non-linearity of the VW functional \cite{alpha}.

We will also consider the exact-XE density (per volume), which is given by:
\begin{equation}
e_{x}(\R)= \sum_{nl,n'l'} e_{x,nl,n'l'} \; 
\end{equation}
{with} \cite{march01,zhou05} 
\begin{widetext}
\begin{eqnarray} 
e_{x,nl,n'l'}(\R) &=& - \sum_{m,m'} \phi_{nlm}^{*}(\R) \phi_{n'l'm'}(\R) \int {\rm d}^3\, \RP
 \frac{\phi_{nlm}^{*}(\RP) \phi_{n'l'm'}(\RP)}{|\R-\RP|} \nonumber \\
     &=& - \frac{1}{4\pi}
        R_{nl}(r) 
        R_{n'l'}(r) 
       \sum_{k} \lambda^k_{l,l'} 
       \int_0^\infty \frac{r_{<}^k}{r_{>}^{k+1}} 
        R_{nl}(r')
        R_{n'l'}(r')
         r'^2 dr' \; 
\end{eqnarray}
where $\lambda^k_{l,l'}=\frac{(2 l+1) (2l'+1)}{2k+1} |\langle l 0 l' 0|k 0 \rangle|^2 $,
$\langle l m l' m'|l''m'' \rangle$ are Clebsch--Gordan coefficients,
$r_<=\min(r,r')$ and $r_>=\max(r,r')$.
In the relevant case of a system with only $s$-shells occupied, we have:
\begin{equation}
e_{x}(\R)     = - \frac{1}{4\pi} \sum_{n,n'} R_{n0}(r) R_{n'0}(r) 
          \int_0^\infty \frac{1}{r_{>}} 
         R_{n0}(r')
         R_{n'0}(r')
         r'^2 dr'\; 
\label{eq:exos}
\end{equation}
\end{widetext}

The above formulas are completely general and apply to any electronic system with a central external potential, e.g., real atoms.
Nevertheless, a very interesting special case is that of hydrogenic orbitals.
In fact, in this case, all calculations are analytical, and explicit formulas can be obtained by any symbolic computer algebra system software.
Moreover, we recall that while in real atoms the electrons far from the nucleus experience
a screened nuclear charge so that the corresponding orbitals differ from the hydrogenic ones, for large atoms or very positive ions, 
this screening effect becomes vanishingly small, and the simple model of hydrogenic orbitals becomes
exact \cite{Lieb}. 
This model system has been largely used in DFT \cite{ioni,march01,march10,march11}, 
is very important for semiclassical physics \cite{LCPB09,burkesemi,ioni} and has been 
used as a main reference system in the {APBE} 
 \cite{apbe} and {APBEk} 
 \cite{apbek} GGA functionals. 

\subsection{Hydrogenic Shells}\par
Contributions near the nucleus are given by $s$-type shells for the KE and the
XE density
and by~$p$-type shells for the KE density only. 
Higher angular momenta do not contribute near the nucleus.

(i) For a filled $s$-shell ($l=0$, for any principal quantum number $n$)
with $f=2$ electrons, we find (after some algebra) near the nucleus (of charge $Z$):
\begin{eqnarray}
\rho_{n0}(r\rightarrow 0) &=& 2\frac{Z^3}{\pi n^3}-2\frac{2Z^4}{\pi n^3}r+\ldots \label{eqs1} \; \\
\tk_{n0} (r\rightarrow 0) &=& 2\frac{Z^5}{2n^3\pi}-2\frac{1}{3}\frac{Z^6(2n^2+1)}{n^5\pi}r+\ldots \; \label{eqs2}\\
\tw_{n0} (r\rightarrow 0)&=& \tk_{n0} (r\rightarrow 0) \; 
\end{eqnarray}
and thus, Kato's theorem \cite{kato}:
\begin{equation}
\tk_{n0}(0) = \tau_{n0}^W (0)= Z^2 \rho(0)/2 \; 
\label{eq4}
\end{equation}
is satisfied for any $n$. 
Equations (\ref{eqs1}) and (\ref{eqs2}) show that at the nucleus, 
all $s$-electrons are important, even if the main contribution
is given by the $n=1$ term (\emph{i.e.}, the $1s$-shell), due to the $n^3$ term at the~denominator.

In case of exchange, we find that the XE density of the shell $n0$ is:
\begin{equation}
e_{x,{n0}}(r\rightarrow 0)=-\frac{Z^4}{\pi n^5} +\frac{2 Z^5}{\pi n^5} r + \ldots \; 
\label{eq5}
\end{equation}

Equation (\ref{eq5}) shows that at the nucleus, only $s$ electrons with very small $n$ contribute, due to the $n^5$ term
at the denominator. {Expression }(\ref{eq5}) generalizes the one in \cite{tao01}. 

For the special case of $n=1$ (\emph{i.e.}, the helium isoelectronic series with fixed hydrogenic orbitals), we~obtain 
the following expressions for the kinetic and exchange enhancement factors: 
\begin{eqnarray}
F_{s}^{HYD1s}(r\rightarrow0)&=&
     \frac{5}{18} \left(\frac{6}{\pi^2}\right)^{1/3} + 
     \frac{10}{27} \left(\frac{6}{\pi^2}\right)^{1/3} Z r + \ldots \nonumber \\
     &\approx&  0.2353 + 0.3138 Z r+ \ldots \; \label{eq11c}\\
F_{x}^{HYD1s}(r\rightarrow0)&=&
     \frac{\left(6\pi\right)^{2/3}}{9} + 
     \frac{2\left(6\pi\right)^{2/3}}{27} Z r + \ldots \nonumber \\
        &\approx& 0.7870 +0.5246 Z r + \ldots \; 
\label{eq6b}
\end{eqnarray}
which can be seen as simple semilocal conditions at the nucleus; see also
\cite{tao01}.
Equation (\ref{eq6b}) shows a significant de-enhancement ($F_x<1$) at the nucleus, which
is not reproduced by conventional DFT functionals (all GGAs and most meta-GGAs).

\par

(ii) For a $p$-shell ($l=1$, for any principal quantum number $n$), we find near the nucleus (of charge $Z$):
\begin{eqnarray}
\rho_{n1}(r\rightarrow 0)&=&2\frac{Z^5(n^2-1)}{3\pi n^5}r^2-2\frac{Z^6(n^2-1)}{3\pi n^5}r^3+\ldots \; \label{eqp1} \\
\tau_{n1}(r\rightarrow 0)&=&2\frac{Z^5(n^2-1)}{2\pi n^5}-2\frac{2Z^6(n^2-1)}{3\pi n^5}r+\ldots \; \label{eqp2} \\
\tau_{n1}^W(r\rightarrow 0)&=&2\frac{Z^5(n^2-1)}{6\pi n^5}-2\frac{Z^6(n^2-1)}{3\pi n^5}r+\ldots \; \label{eqp3} 
\label{eqp5}
\end{eqnarray}

For a system with only the $p$-shell occupied, there is no cusp of the density, and Kato's formula cannot be applied.
Interestingly and importantly, even if $\rho_{n1}(0)=0$ at the nucleus, the kinetic energies are
of the same order of magnitude ($\sim Z^5$) as in the $s$-shell case. 
From Equations (\ref{eqp2}) and (\ref{eqp3}), we find:
\begin{equation}
\tau_{n1}(0)/\tau_{n1}^W(0)=3 \; 
\label{eq8}
\end{equation}
for any principal quantum number, in agreement with Equation (\ref{eq1}).

Similarly, the XE density at the nucleus, for the $p$-shells, is: 
\begin{equation}
e_{x,n1}(r\rightarrow 0)=-\frac{1}{3}\frac{(n+1)Z^6}{\pi n^8}r^2 + \ldots \; 
\label{eq9}
\end{equation}
and thus, for the exchange case, the $p$-orbitals do not contribute at the nucleus.

\subsection{Ten-Electron Hydrogenic Model}
\label{sec:ten}
Now, let us consider a 10-electron hydrogenic atom, with electronic structure $1s^22s^22p^6$ and nuclear charge $Z$
(\emph{i.e.}, the Ne isoelectronic series with fixed hydrogenic orbitals). 
The calculations are analytical, and we obtain (after some algebra):
\begin{eqnarray}
\rho(r\rightarrow 0) &=& 2\frac{9Z^3}{8\pi} -2\frac{9Z^4}{4 \pi }r + \ldots \; \\
\tk(r\rightarrow 0) &=& 2\frac{39 Z^5}{64\pi} -2\frac{37 Z^6}{32 \pi }r + \ldots \; \label{eq:10k}\\
\tw(r\rightarrow 0) &=& 2\frac{9 Z^5}{16\pi} -2\frac{9 Z^6}{8 \pi }r + \ldots \; \\
F_s(r\rightarrow 0)&=&\frac{65}{162} \left(\frac{2}{\pi^2}\right)^{1/3} + 
          \frac{140}{243} \left(\frac{2}{\pi^2}\right)^{1/3} Z r + \ldots \nonumber \\
         &\approx&0.2357 + 0.3384 Z r+\ldots \; \label{eq11a}\\ 
F_s^{W}(r\rightarrow 0)&=& \frac{10}{27} \left (\frac{2}{\pi^2} \right )^{1/3} + 
          \frac{40}{81} \left(\frac{2}{\pi^2}\right)^{1/3} Z r + \ldots \nonumber \\
          &\approx& 0.2175 + 0.2901 Z r + \ldots \; \label{eq11b}
\end{eqnarray}

Note that Equation (\ref{eq:10k}) confirms the validity of Equation (\ref{eq1}), and
Equations (\ref{eq11a}) and (\ref{eq11b}) confirm that
the VW approximation underestimates the exact result at the nuclear cusp and in a region close to the nucleus
(the coefficients of $Z r$ are different).

In the case of exchange, we obtain:
\begin{eqnarray}
e_{x}(r\rightarrow 0)&=&-\frac{1019}{864} \frac{Z^4}{\pi} + \frac{1019}{432} \frac{Z^5}{\pi}r + \ldots \, \label{exx10} \\
F_x(r\rightarrow 0)&=& \frac{1019}{4374}(2\pi)^{2/3}+\frac{1019}{6561}(2\pi)^{2/3} Z r + \ldots \nonumber \\
        &\approx&0.7932+0.5288 Zr + \ldots \; \label{fxx10}
\end{eqnarray}

Note that the two leading terms in Equations (\ref{exx10}) and (\ref{fxx10}) are the same in the case
of a Be isoelectronic series, since the $p$-shell contributes to the exchange only with power $r^2$ or higher. 

\subsection{The Asymptotic Neutral Atom with an Infinite Number of Electrons}
Using Equations (\ref{eq:rs}) and (\ref{eq:ts})
and the Riemann $\zeta$-function 
(\emph{i.e.}, $\zeta(s)=\sum_{n=1}^\infty 1/n^s$), 
we obtain the following analytical expressions near the nucleus:
\begin{eqnarray}
\rho(r\rightarrow 0) &=&2 \frac{Z^3\zeta(3)}{\pi} - \frac{2Z^4\zeta(3)}{\pi} r + \ldots \nonumber \\ 
         &=& 0.7653 Z^3-1.5305 Z^4 r + \ldots \; \\
\tk(r\rightarrow 0) &=&2 \left( \frac{ \zeta(3)}{ \pi} 
             - \frac{ \zeta(5)}{2\pi} \right) Z^5 - \\
\nonumber
&& -           2\left( \frac{4\zeta(3)}{3\pi} 
             - \frac{ \zeta(5)}{3\pi} \right) Z^6 r + \ldots \nonumber \\
        &=& 0.4352 Z^5 - 0.8003 Z^6 r + \ldots \; \\
\tau^W(r\rightarrow 0) &=& 2 \frac{Z^5\zeta(3)}{2\pi} - 2\frac{Z^6\zeta(3)}{\pi} r + \ldots \nonumber \\ 
         &=& 0.3826 Z^5-0.7652 Z^6 r + \ldots \; 
\label{eq11bis}
\end{eqnarray}
\begin{eqnarray}
\nonumber
F_s (r\rightarrow 0)&=& \frac{5}{18} \left( \frac{6}{\pi^2 } \right)^{1/3} \frac{2\zeta(3)-\zeta(5)}{\zeta(3)^{5/3}} + \\
\nonumber
         &&  +\frac{10}{27} \left( \frac{6}{\pi^2 } \right)^{1/3} \frac{2\zeta(3)-\zeta(5)}{\zeta(3)^{5/3}} Zr + \ldots \nonumber \\
          &\approx& 0.2367+0.3538 Z r+ \ldots \; \label{fszi}\\
F_s^{W}(r\rightarrow 0) &=& \frac{5}{18} \left( \frac{6}{\pi^2 \zeta(3)^2} \right)^{1/3}
             +\frac{10}{27} \left( \frac{6}{\pi^2 \zeta(3)^2} \right)^{1/3} Z r + \ldots \nonumber \\
           &=& 0.2081+0.2775 Z r + \ldots \; 
\end{eqnarray}

As expected, the difference between the exact KE density and the VW one
is larger than in the case of ten electrons:
interestingly, the coefficients of $Z r$ differ more significantly than the values in $r=0$.
Note that at the nuclear cusp of an atom with $Z\rightarrow \infty$ electrons,
the reduced gradient is $s=0.3534$, \emph{i.e.}, smaller than for the helium isoelectronic series ($s=0.375$).

For the exchange case, no analytic results could be found. In fact, 
even if only the $s$-type shell contributes to the nuclear cusp, Equation (\ref{eq:exos}) involves double sums, and
we could not find a~closed form expression for the terms with $n \neq n'$. 
We performed the double sums numerically up to $n=n'=50$ obtaining: 
\begin{equation}
F_x(0)\approx 0.798 \; \label{fxinf}
\end{equation}

\subsection{1s-Shell Model} 
Comparing the results of the previous three sections, we note that 
the exact KE enhancement factors $F_s$ are very similar in all three cases (see Equations (\ref{eq11c}), (\ref{eq11a}) and (\ref{fszi})), 
with a maximum deviation of only $0.0014$.
This is not the case for $F_s^W$ (deviation 20-times larger).
This means that the KE enhancement factor near the nuclear cusp of an atom with an infinite number of electrons
 is almost equivalent to the one obtained from the $1s$ shell only.

We can thus estimate the KE density near the nuclear cusp via a simple procedure, which we call the $1s$-shell model (1SM): 
for a given atom, we consider only the 
density form the $1s$-shell and use it to compute the KE density (which then equals the VW one), \emph{i.e.},
\begin{equation}
\tau^{1SM}[\rho](\R)=\tau^W[\rho_{1s}](\R) \; 
\label{eq1sm}
\end{equation}

For the neon hydrogenic isoelectronic series, the 1SM gives very accurate results: this, however, traces back
to a subtle error cancellation between the $2s$ and $2p$ contributions.

It is important to underline that, despite the simple expression in
Equation (\ref{eq1sm}), the $\rho_{1s}$ can hardly be described by any semilocal 
ingredient of the total density $\rho$: 
in other words, the mapping $\rho \rightarrow \rho_{1s}$ is highly non-local.
We can conclude that $F_s\approx 0.235$ at the nuclear cusp will yield accurate results for all atoms, from He to the semiclassical
limit, whereas $F_s\approx F_s^W$ is not accurate.

Concerning the exchange, comparing Equations (\ref{eq6b}), (\ref{fxx10}) and (\ref{fxinf}), we see that all of the exact exchange enhancement factors $F_x$ 
are similar to each other.

Thus, also for the exchange case, we can define the 1SM approach as:
\begin{equation}
e_x^{1SM}[\rho](\R)=e_x[\rho_{1s}](\R)=\frac{1}{4}\rho_{1s}(\R)\int {\rm d}^3\RP\, \frac{\rho_{1s}(\RP)}{|\R-\RP|} \; 
\label{eq1km}
\end{equation}

A similar approach has been used to compute the exchange potential at the nucleus \cite{liu95}.

In the next section, we will show that the 1SM model will yield good accuracy for both the KE and XE density,
also in the case of real atoms.

\subsection{Real Atoms} 
\label{sec:ra}
We considered non-relativistic neutral noble atoms using 
self-consistent numerical Kohn--Sham {
 exact exchange orbitals and densities \cite{ioni,CP3}. 
The wavefunctions $R_{nl}$ are discretized on a semi-logarithmic numerical} grid, using the Engel code 
\cite{engel1,engel2}.

\begin{figure}
\begin{center}
\includegraphics[width=0.9\columnwidth]{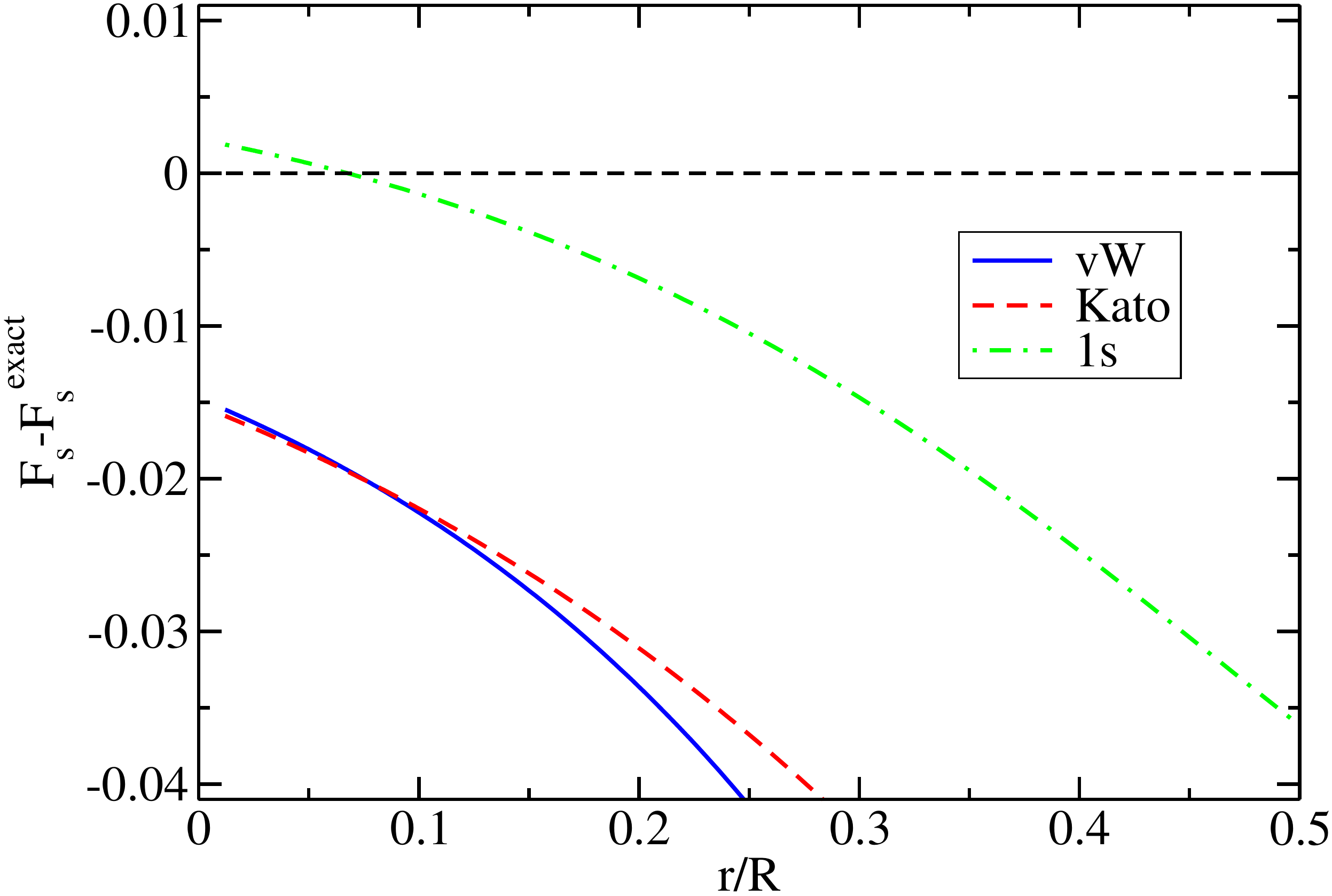}
\caption{Error on the kinetic enhancement factor $F_s-F_s^{exact}$ \emph{versus} the
scaled radial distance $r/R$ near the nucleus of the Kr atom.
\label{f1}}
\end{center}
\end{figure}

In Figure \ref{f1}, we show the error on the kinetic enhancement factor $F_s-F_s^{exact}$ \textit{versus} the 
scaled radial distance $r/R$ near the nucleus of the Kr atom: we consider 
the VW functional, the 
Kato approximation (\emph{i.e.}, $\tau^{Kato}=Z^2\rho/2$) 
and the 1SM method.
The scaled radial distance is defined as the average distance of the $1s$ shell:
\begin{equation}
R=\frac{1}{2}\int {\rm d}^3\R\, \rho_{1s}(\R) r \; 
\label{eq12bb}
\end{equation}
where $\rho_{1s}$ is the density of the $1s$ shell 
(and $\int d\mathbf{r} \rho_{1s}=f=2$). 
For real atoms, we find that $R\approx 3/(2Z)$.

Figure \ref{f1} 
shows that:
(i) the VW approximation is not exact at the nuclear cusp; (ii) the 
Kato expression is a very good model for the VW behavior, not only
at the nuclear cusp, but also for $r/R < 0.2$ \cite{spatgen};
(iii) the 1SM 
is accurate at the nucleus.

\begin{figure}
\begin{center}
\includegraphics[width=0.9\columnwidth]{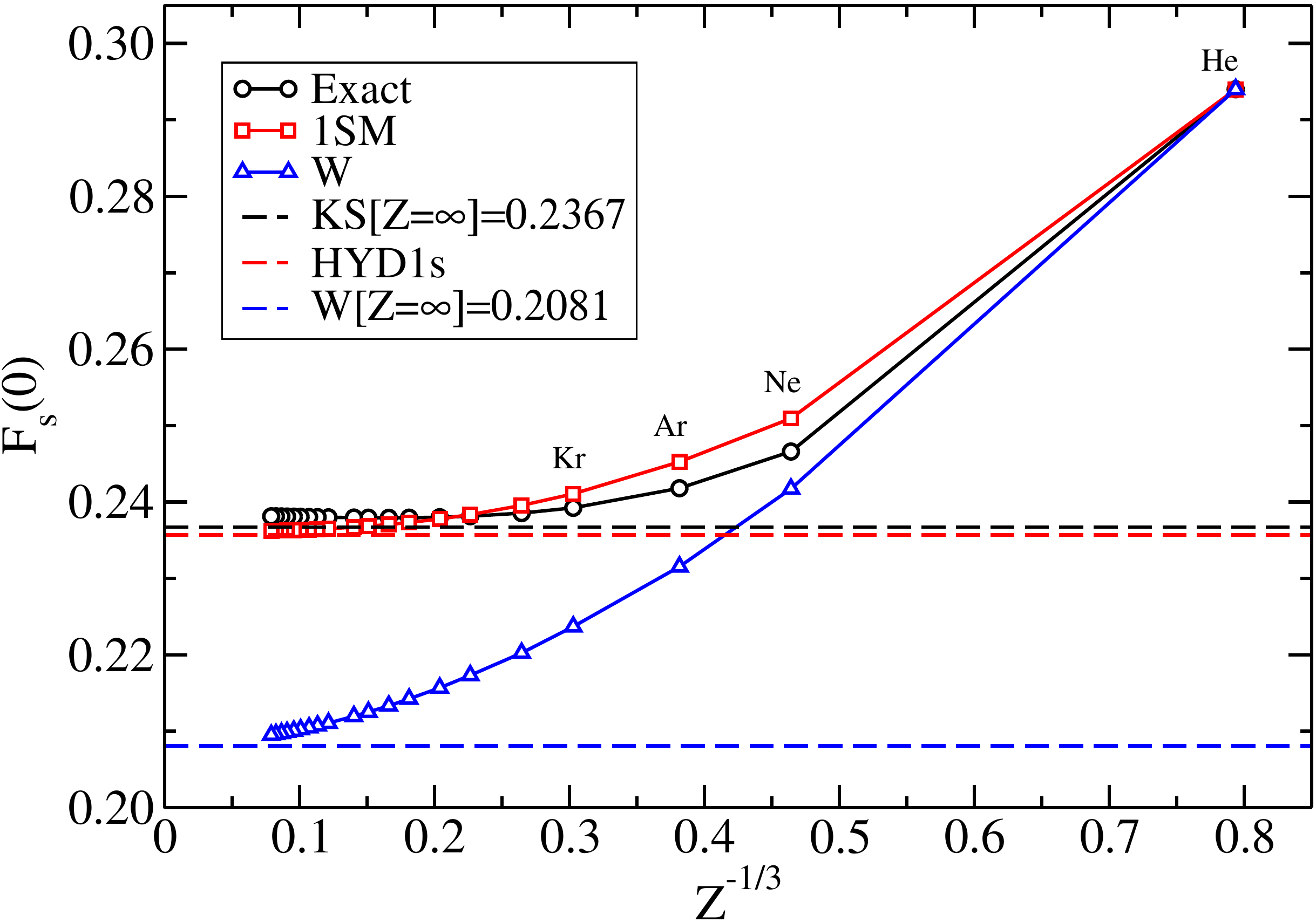}
\caption{Kinetic energy enhancement factor $F_s$ at the nuclear cusp, for the noble atoms ($2\leq Z\leq 2022$); see the text for details.
\label{f1s}}
\end{center}
\end{figure}

Figure \ref{f1s} reports $F_s$ at the nuclear cusp, for neutral noble atoms with filled shells with the number of electrons in the 
range $2\leq Z\leq 2022$ \cite{alpha}, 
considering the exact $F_s$, the VW functional and the 1SM approach.
Note that the value $F_s(r=0)$ has been extrapolated from the available numerical grid point closest to the nucleus. 
For the He atom, all approaches coincide ($F_s\approx 0.3$): the results differ from the $F_s^{HYD1s}$ of Equation (\ref{eq11c}) due to the screening 
effects, which are largest for the He atom, but rapidly decrease with increasing $Z$.
For larger atoms, the exact $F_s$ converges to 0.2367 (see Equation~(\ref{fszi})), while~$F_s^W$ is much lower.
On the other hand, the 1SM method nicely reproduces the exact results for almost all systems, converging to the HYD1s
 value 
(see Equation \ref{eq11c}) for $Z\rightarrow \infty$.

\begin{figure}
\begin{center}
\includegraphics[width=0.9\columnwidth]{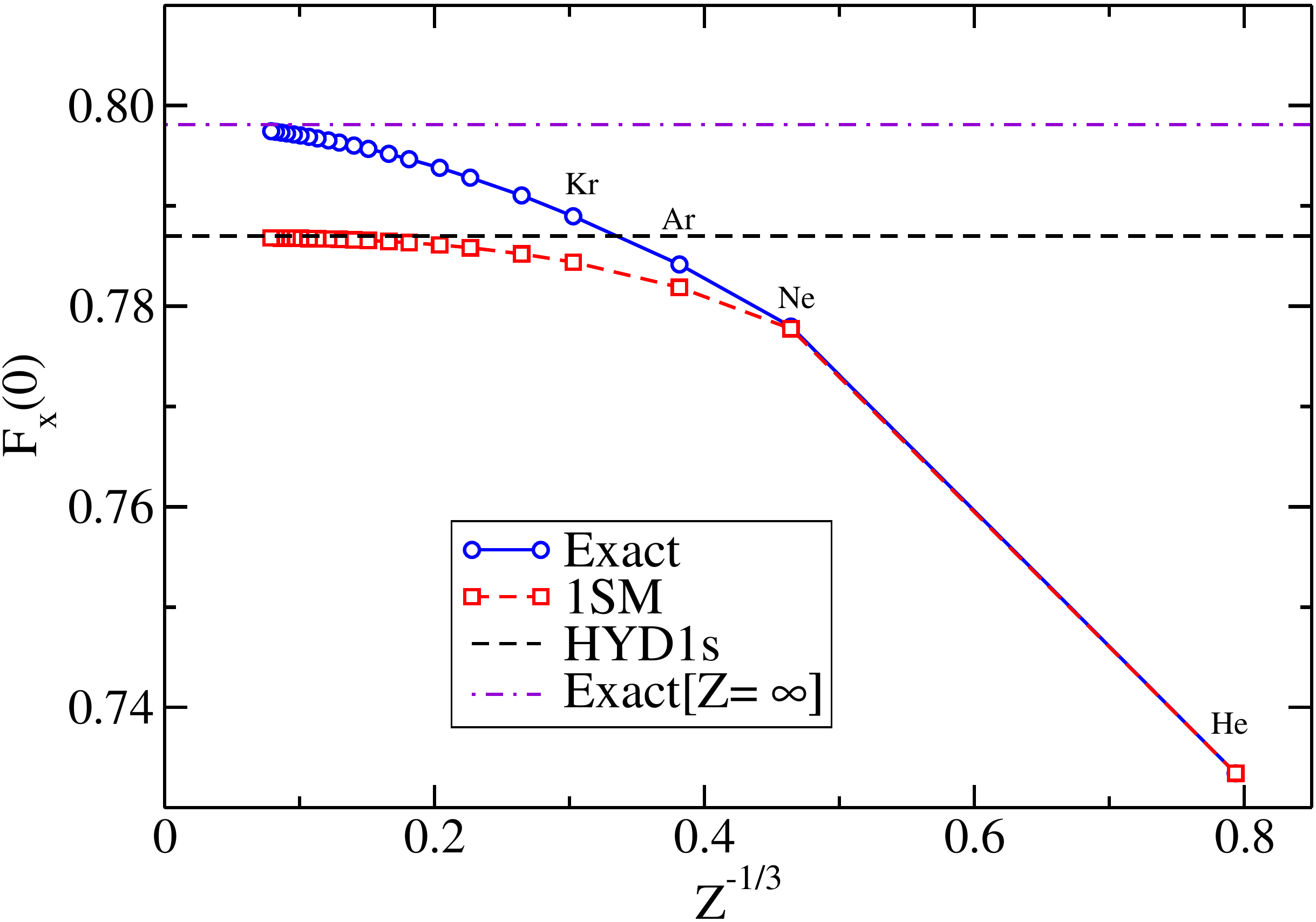}
\caption{Exchange enhancement factors $F_x$ at the nucleus, for the noble atoms ($2\leq Z\leq 2022$); see the text for details.
\label{f2}}
\end{center}
\end{figure}

We now turn to the exchange case.
In Figure \ref{f2}, we show the exact exchange enhancement factor at the nucleus, the 1SM approach and the 
$HYD1s$ value of Equation (\ref{eq6b}), 
for noble atoms ($2\leq Z\leq 2022$). 
The 1SM approach yields very accurate results only for the smallest atoms (up to Ar). 
For the largest atoms, the differences are significantly larger than in the kinetic case: 
in fact, the XE density is more non-local than the KE one.
The simple expression $F_x^{HYD1s}(0)=0.787$ is also accurate and can be 
used for the construction of more realistic semilocal exchange functionals \cite{tao01}.

\section{Non-Local Approximations for Exchange and Kinetic Energies, at and near the Nucleus}\label{sec:nl}
As shown above, the kinetic and exchange energy densities are fully non-local near the nucleus, and thus, 
their behaviors cannot be well captured by semilocal ingredients.

\begin{figure}
\begin{center}\includegraphics[width=0.9\columnwidth]{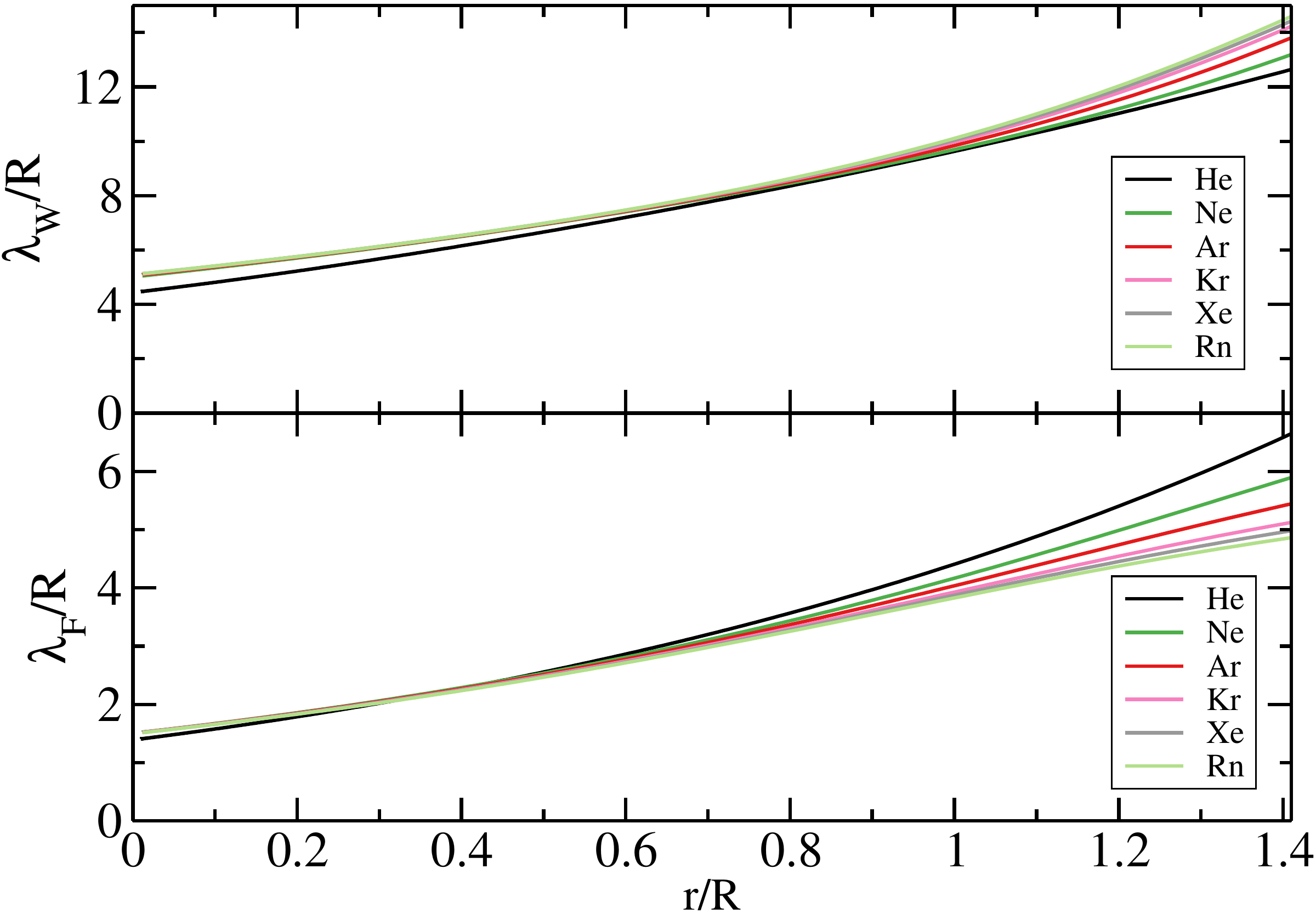}
\caption{The scaled lengths $\lambda_F/R$ and $\lambda_W/R$, \textit{versus} the scaled distance $r/R$, for noble 
atoms (He-Rn). \label{f2bis}
}
\end{center}
\end{figure}

In order to build non-local ingredients to describe better the features of the exchange and kinetic functionals, first we need to consider 
appropriate lengths. In Figure \ref{f2bis}, we compare the VW and the~Fermi wavelengths, defined by: 
\begin{eqnarray} 
\lambda_W&=&2\pi/(\tau^W)^{1/5} \; \label{eq:www}\\
\lambda_F&=&2\pi/k_F=2\pi/(3\pi^2\rho)^{1/3} \label{eq:fww}\; 
\end{eqnarray}

The VW length is similar for all atoms that contain $p$-orbitals 
(Ne-Rn) and is slightly different for the He atom. Thus, $\lambda_W$ can distinguish between atoms that contain 
only $s$-orbitals (e.g., He) and the other atoms (e.g., Ne-Rn). For this reason, we chose this length for the 
kinetic case.

\begin{figure}
\begin{center}
\includegraphics[width=0.9\columnwidth]{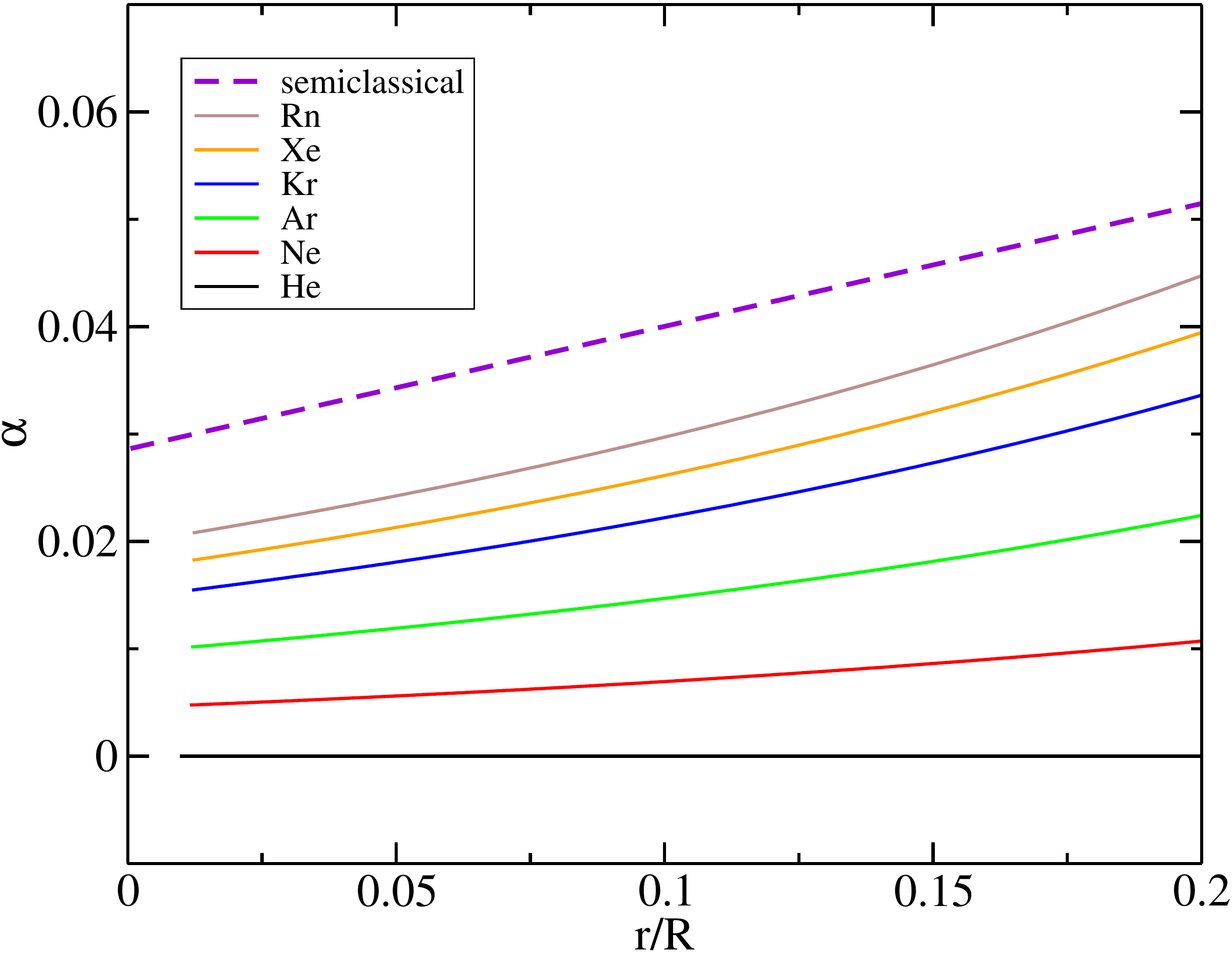}
\caption{$\alpha$ \textit{versus} the scaled radial distance $r/R$, near the nucleus of noble atoms.}
\label{f4}
\end{center}
\end{figure}

On the other hand, the Fermi wavelength near the nucleus is similar for all of the atoms (He-Rn) reported in the figure.
We will use this length, combined with the other meta-GGA ingredient $\alpha=(\tk-\tw)/\tau^\HEG=F_s-F_s^W$, for the exchange 
case.
In Figure \ref{f4}, we report $\alpha$
\textit{versus} the scaled radial distance $r/R$, near the nucleus of 
noble atoms, together with the semiclassical asymptotic limit: 
\begin{eqnarray}
\alpha(r\rightarrow 0)&=&
 \frac{5}{18} \left( \frac{6}{\pi^2}\right)^{1/3}\frac{\zeta(3)-\zeta(5)}{\zeta(3)^{5/3}}+ \nonumber \\
&&\frac{20}{27} \left( \frac{6}{\pi^2}\right)^{1/3}\frac{\zeta(3)-\zeta(5)}{\zeta(3)^{5/3}} Z r + \ldots \nonumber \\ 
&\approx& 0.0286+0.076Zr+\ldots \; 
\label{eq12b3}
\end{eqnarray} 
which shows that $\alpha$ is not vanishing at the nucleus. 
Figure \ref{f4} shows that $\alpha$ has a monotonic behavior for the noble atoms series, 
starting from $\alpha=0$ in the case of the He atom, while for the Rn atom, it becomes quite close to the asymptotic limit.  
We will use this finding in order to construct a proper ingredient for exchange.
\unskip

 \subsection{Kinetic Energy}
In this section, we build a new non-local density ingredient suitable for 
the description of the KE density near the nucleus, \emph{i.e.}, not only at the nuclear cusp, but also in a region around the nucleus.

We start from the observation that the 1SM model of Equation (\ref{eq1sm}) is almost exact; see Figure \ref{f1s}.
Then, instead of considering the VW functional of the density of the $1s$-shell,
we consider a screened VW functional of the total density.

Thus, we propose the following $\beta$-family of ingredients:
\begin{equation}
y^W(\R)=\frac{a}{\rho(\R)^{2/3}}\int {\rm d}^3 \R' \tau^W(\R') e^{-b [\tau^W(\R')]^\beta|\R-\R'|^{5\beta}} \; 
\label{eq14}
\end{equation}
where $a$, $b$ and $\beta$ are positive constants. 
The non-local ingredient $y^W$ is invariant under the uniform scaling 
of the density (as $s$ and $q$), and thus, it can be used in the construction of KE enhancement factor 
approximations. For simplicity, in this paper, we chose $\beta=1$. 
Note that the exponential damping factor of Equation (\ref{eq14}) 
uses the VW length of Equation (\ref{eq:www}). 

The parameters $a$ and $b$ have been optimized for the Ne atom, in the region $r/R\leq 1$, by~minimizing the following error:
\begin{equation}
Error=\int^R_0 dr\; |F_s^{exact}(r)-y^W(r)| \; 
\end{equation} 
finding $a=0.2$ and $b=0.08$, that completely defines the $y^W$ ingredient.

\begin{figure}
\begin{center}
\includegraphics[width=0.9\columnwidth]{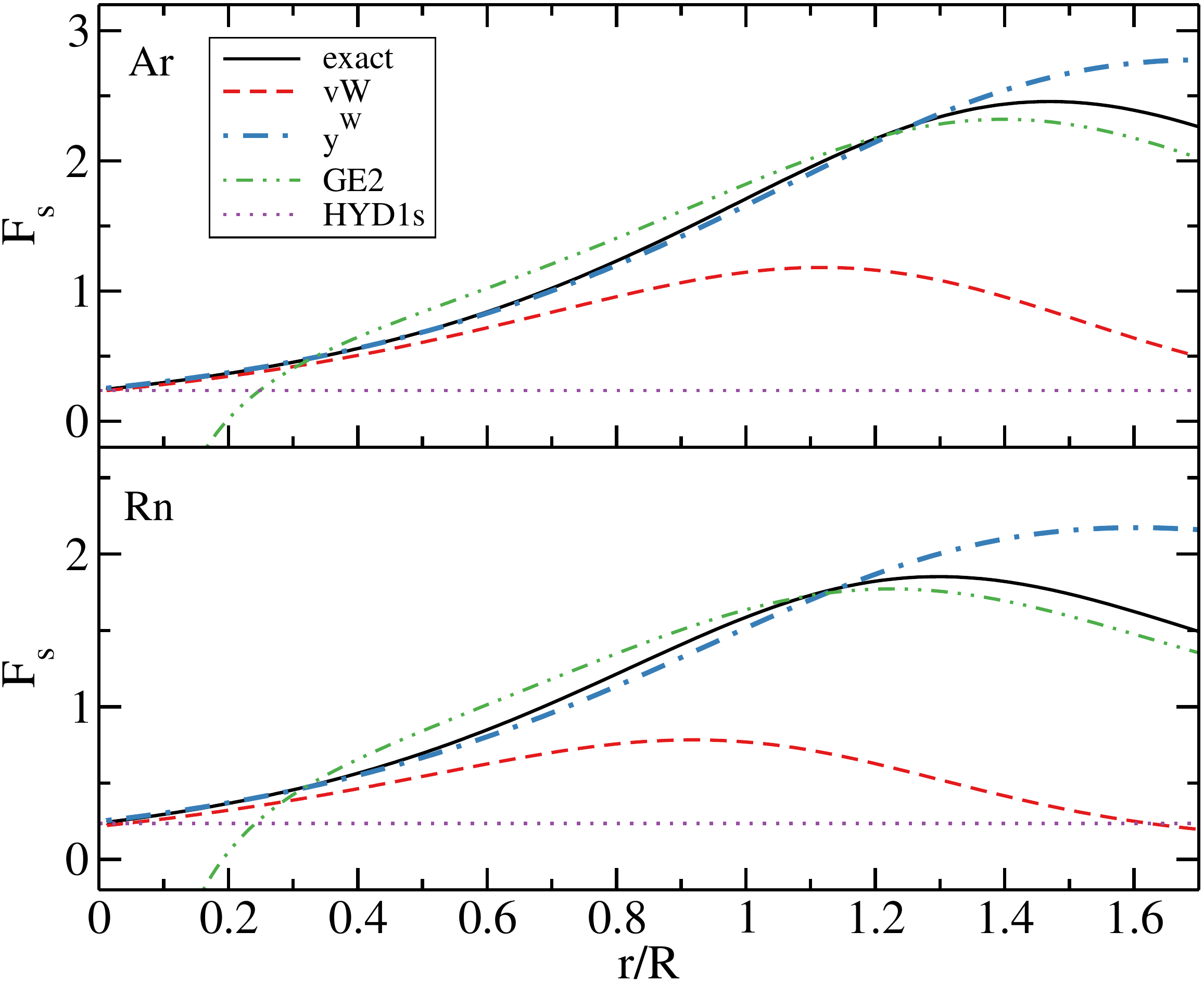}
\caption{Kinetic energy enhancement factors $F_s$ \textit{versus} scaled radial distance, for the Ar atom (upper panel) and the Rn atom (lower panel). 
\label{f3}}
\end{center}
\end{figure}

In Figure \ref{f3}, we report a comparison between the exact KE enhancement factor, the VW 
one and~the~$y^W$, for Ar and Rn atoms. The ingredient $y^W$ is accurate near the nucleus, 
while the second-order gradient expansion (GE2) enhancement factor ($F_s^{GE2}=1+5s^2/27+20q/9$) diverges 
at the nucleus (as $q\rightarrow -\infty$).
Similar good results have been found for other noble gas atoms ({data not reported}). 
Note that $y^W$ is not a good model for $r/R>1.2$, where instead, GE2 is closer to the exact kinetic
enhancement factor.

The non-local ingredient $y^W$ opens the possibility of constructing KE approximations of the form:
\begin{equation}
T_s=
\int {\rm d}^3\R \; \tau^\HEG F_s(s,q,y^W) \; 
\label{eq15}
\end{equation}
and the functional derivative of such an expression can be computed starting from its definition. After~some algebra, 
we obtain: 
\begin{eqnarray}
& \frac{\delta T_s}{\delta 
\rho}=\frac{\partial\tau}{\partial\rho}-\nabla(\frac{\partial\tau}{\partial\nabla\rho})+
\nabla^2(\frac{\partial\tau}{\partial\nabla^2\rho})-
\frac{2y^W}{3\rho}\frac{\partial\tau}{\partial y^W}+\nonumber\\
& \int 
{\rm d}^3\R'\frac{\partial\tau}{\partial y^W}(\R')
\frac{a}{\rho^{2/3}(\R')}e^{-b\tau^W(\R)|\R-\R'|^5}f(\R,\R') \; 
\label{eq16}
\end{eqnarray}
with: 
\begin{eqnarray}
&& f(\R,\R')=\frac{b^2 |\nabla\rho(\R)|^4|\R-\R'|^{10}}{32\rho^2(\R)}\frac{\delta 
T^W}{\delta\rho(\R)}- \nonumber\\
&& \frac{5b |\nabla\rho(\R)|^2 |\R-\R'|^{3} (\R-\R')\cdot\nabla \rho(\R)}{16\rho^2(\R)}
 (-1+\frac{b}{2}|\R-\R'|^5 \tau^W(\R))-\nonumber\\
&& \frac{b |\nabla\rho(\R)|^2 |\R-\R'|^{5}}{32\rho^2(\R)}
(20\tau^W(\R)-5\nabla^2\rho(\R))+\frac{\delta T^W}{\delta\rho(\R)}\; 
\label{eq17} 
\end{eqnarray}

Equation (\ref{eq16}) differs from a semilocal expression mainly due to an extra integral, but it is still 
numerically feasible, having the same computational cost as the total KE $T_s$.

\subsection{Exchange Energy}
%
%
\begin{figure}
\begin{center}
\includegraphics[width=0.9\columnwidth]{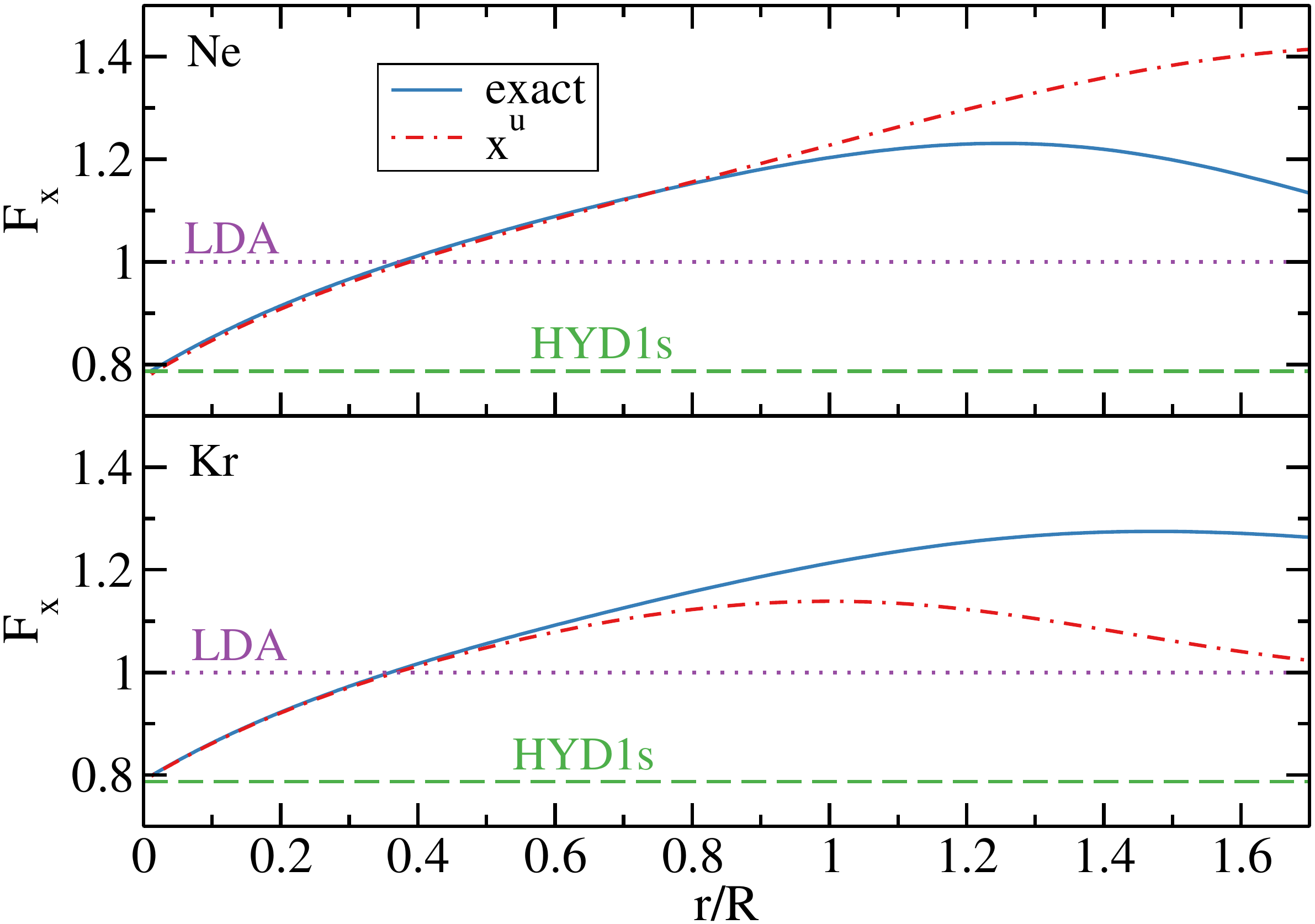}
\caption{{Exchange enhancement factors $F_x$ near the nucleus \textit{versus} scaled radial distance, for the Ne atom (upper panel) and the Kr atom (lower panel)}. 
\label{f5}
}
\end{center}
\end{figure} 
%

For the exchange case, we start again from the observation that the 
1SM model of Equation (\ref{eq1km}) is almost exact; see Figure \ref{f2}.
Then, instead of considering the Hartree potential of the density of the $1s$-shell,
we consider a screened Hartree potential of the total density.

Thus, in order to describe XE density in a region near the nucleus, 
we introduce a $\beta$-family of ingredients, which is invariant under the uniform density scaling, of the form:
\begin{equation}
x^u(\R)=\frac{1}{3(3\rho(\R)/\pi)^{1/3}}\int {\rm d}^3\R' \frac{\rho(\R')}{|\R-\R'|}e^{-a 
\alpha(\R')^b k_F(\R')^\beta|\R-\R'|^{\beta}} \;
\label{eq18}
\end{equation}
where $a$, $b$ and $\beta$ are positive constant.
Note that the exponential dumping factor of Equation (\ref{eq18})
uses the Fermi wavelength of Equation (\ref{eq:fww}).
Near the nucleus of a 
many-electron atom, $\alpha$ is small, but nonzero, being a measure of the $p$-electrons contribution.

We recall that for any one- and two-electron systems, $\alpha=0$, the 1SM model of Equation (\ref{eq1km}) is exact, and thus,
$F_x^{exact}=x^u$.
Clearly, an exact $F_x$ for one- and two-electron systems is a 
very important condition in DFT that cannot be reached by any semilocal functional. 
In this respect, the here proposed 
ingredient $x^u(\R)$ can be useful in further developments of non-local functionals.

In Equation (\ref{eq18}), for simplicity, we chose $\beta=1$, and the parameters $a$ and $b$ have been 
optimized for the Ne atom, using the same procedure as in the kinetic case.
We find that for $a=0.8$, $b=0.85$, $x^u(\R)$ is suitable for the description of 
the nuclear region up to $r/R\approx0.6-0.8$, as shown in Figure \ref{f5}, in the cases of Ne and Kr atoms. 

As in the KE case, non-local XE functionals with improved nuclear behavior can be developed, 
having a general expression (for spin-unpolarized systems):
\begin{equation}
E_{x}=
\int {\rm d}^3\R\, e_{x}^\HEG F_x(s,q,\alpha,x^u) \; 
\label{eq19}
\end{equation}
and using the generalized Kohn--Sham method \cite{seidl96}, the XE potential can be computed in the same manner, 
as has been shown in Equation (\ref{eq16}), and will differ from a regular meta-GGA potential only due 
to an extra integration.
Finally, we remark that the exact nuclear behavior can be reproduced only by~high level 
functionals, which include the exact-XE density, such as the 
optimized-effective potential (OEP) method \cite{talman76,kummelrev,lhf} or 
hyper-GGAs methods \cite{hyper1,hyper2,hyper3}, which are significantly more 
expensive than the proposed Equation (\ref{eq19}).

\section{Conclusions} \label{sec:cl}

In conclusion, we have investigated the behaviors of the kinetic and exchange energies near the nucleus region. 
By employing the simple, but very powerful hydrogenic orbital model system, we~have reported the exact expression for
the kinetic and exchange enhancement factor near the~nucleus, from~the~helium isoelectronic series to the semiclassical
limit of a neutral atom with an infinite number of electrons.
This analytical study has also proven that the $1s$-model is very accurate for the kinetic 
(due to a subtle error compensation mainly between the $2s$ and $2p$ electrons) and for the exchange energy density. 

The physics of the kinetic and exchange energy densities near the nucleus region has fully non-local features, and thus, it cannot be 
captured by the usual semilocal ingredients. For this reason, semilocal exchange-correlation and kinetic
 functional approximations are not accurate in this region. We propose density-dependent ingredients 
(\emph{i.e.}, $y^W$ of Equation (\ref{eq14}) and $x^u$ of Equation (\ref{eq18})) that can 
well describe this important density region and can become attractive tools for the future development of DFT functionals.


\begin{thebibliography}{999}

\bibitem[Kohn and Sham(1965)]{KS}
Kohn, W.; Sham, L.J.
\newblock Self-consistent equations including exchange and correlation effects.
\newblock {\em Phys. Rev.} {\bf 1965}, {\em 140},~A1133--A1138.

\bibitem[Dreizler and Gross(1990)]{dreizbook}
Dreizler, R.M.; Gross, E.K.U.
\newblock {\em Density Functional Theory}; Springer: New York, NY, USA, 1990.

\bibitem[Parr and Yang(1989)]{parrbook}
Parr, R.G.; Yang, W.
\newblock {\em Density-Functional Theory of Atoms and Molecules}; Oxford
 University Press: Oxford, \linebreak UK, 1989.

\bibitem[Scuseria and Staroverov(2005)]{scusrev}
Scuseria, G.E.; Staroverov, V.N.
\newblock Progress in the development of exchange-correlation functionals. In
 {\em Theory and Applications of Computational Chemistry: The First 40 Years
 (A Volume of Technical and Historical Perspectives)}; Dykstra, C.E.,
 Frenking, G., Kim, K.S., Scuseria, G.E., Eds.; Elsevier: Amsterdam, The Netherlands, 2005;\linebreak 
 pp. 669--724.

\bibitem[Becke(2014)]{becke14}
Becke, A.D.
\newblock Perspective: Fifty years of density-functional theory in chemical
 physics.
\newblock {\em J. Chem. Phys.} {\bf 2014}, {\em 140}, doi:10.1063/1.4869598.

\bibitem[Peverati and Truhlar(2014)]{peveratirev}
Peverati, R.; Truhlar, D.G.
\newblock Quest for a universal density functional: The accuracy of density
 functionals across a broad spectrum of databases in chemistry and physics.
\newblock {\em Phil. Trans. A} {\bf 2014}, {\em 372},~doi:10.1098/rsta.2012.0476.

\bibitem[Wesolowski \em{et~al.}(2015)Wesolowski, Shedge, and Zhou]{wesochemrev}
Wesolowski, T.A.; Shedge, S.; Zhou, X.
\newblock Frozen-density embedding strategy for multilevel simulations of
 electronic structure.
\newblock {\em Chem. Rev.} {\bf 2015}, {\em 115},~5891--5928.

\bibitem[Jacob and Neugebauer(2014)]{jacobrev}
Jacob, C.R.; Neugebauer, J.
\newblock Subsystem density-functional theory.
\newblock {\em WIRE} {\bf 2014}, {\em 4},~325--362.

\bibitem[Krishtal \em{et~al.}(2015)Krishtal, Sinha, Genova, and
 Pavanello]{pavanello15}
Krishtal, A.; Sinha, D.; Genova, A.; Pavanello, M.
\newblock Subsystem density-functional theory as an effective tool for modeling
 ground and excited states, their dynamics and many-body interactions.
\newblock {\em J. Phys. Condens. Matter} {\bf 2015}, {\em 27},~183202.

\bibitem[Wang and Carter(2000)]{wangcarterof}
Wang, Y.; Carter, E.A.
\newblock Orbital-free kinetic-energy density functional theory. In {\em
 Progress in Theoretical Chemistry and Physics}; Schwartz, S., Ed.; Kluwer:
 Dordrecht, The Netherlands, 2000; p. 117.

\bibitem[Wesolowski and Wang(2013)]{wesobook}
Wesolowsky, T.A.; Wang, Y.A.
\newblock {\em Recent Progress in Orbital-Free Density Functional Theory};
 World Scientific: Singapore, 2013.

\bibitem[Xia \em{et~al.}(2012)Xia, Huang, Shin, and Carter]{xia12}
Xia, J.; Huang, C.; Shin, I.; Carter, E.A.
\newblock Can orbital-free density functional theory simulate molecules?\linebreak
\newblock {\em J. Chem. Phys.} {\bf 2012}, {\em 136},~doi:10.1063/1.3685604.

\bibitem[Karasiev and Trickey(2012)]{karasiev12}
Karasiev, V.; Trickey, S.
\newblock Issues and challenges in orbital-free density functional
 calculations.
\newblock {\em Comput. Phys.~Commun.} {\bf 2012}, {\em 183},~2519--2527.

\bibitem[Perdew \em{et~al.}(1996)Perdew, Burke, and Ernzerhof]{pbe}
Perdew, J.P.; Burke, K.; Ernzerhof, M.
\newblock Generalized gradient approximation made simple.
\newblock {\em Phys. Rev. Lett.} {\bf 1996}, {\em 77},~3865--3868.

\bibitem[Becke(1988)]{b88}
Becke, A.D.
\newblock Density-functional exchange-energy approximation with correct
 asymptotic behavior.
\newblock {\em Phys.~Rev.~A} {\bf 1988}, {\em 38},~3098--3100.

\bibitem[Zhang and Yang(1998)]{revpbe}
Zhang, Y.; Yang, W.
\newblock Comment on ``Generalized gradient approximation made simple''.
\newblock {\em Phys. Rev. Lett.} {\bf 1998}, {\em 80},~doi:10.1103/PhysRevLett.80.890.

\bibitem[Hammer \em{et~al.}(1999)Hammer, Hansen, and N\o{}rskov]{rpbe}
Hammer, B.; Hansen, L.B.; N\o{}rskov, J.K.
\newblock Improved adsorption energetics within density-functional theory using
 revised Perdew-Burke-Ernzerhof functionals.
\newblock {\em Phys. Rev. B} {\bf 1999}, {\em 59},~7413--7421.

\bibitem[Wu and Cohen(2006)]{wc}
Wu, Z.; Cohen, R.E.
\newblock More accurate generalized gradient approximation for solids.
\newblock {\em Phys. Rev. B} {\bf 2006}, {\em 73}, doi:10.1103/PhysRevB.73.235116.

\bibitem[Haas \em{et~al.}(2011)Haas, Tran, Blaha, and Schwarz]{htbs}
Haas, P.; Tran, F.; Blaha, P.; Schwarz, K.
\newblock Construction of an optimal GGA functional for molecules and solids.
\newblock {\em Phys. Rev. B} {\bf 2011}, {\em 83},~doi:10.1103/PhysRevB.83.205117.

\bibitem[Constantin \em{et~al.}(2011)Constantin, Fabiano, Laricchia, and {Della
 Sala}]{apbe}
Constantin, L.A.; Fabiano, E.; Laricchia, S.; {Della Sala}, F.
\newblock Semiclassical Neutral Atom as a Reference System in Density
 Functional Theory.
\newblock {\em Phys. Rev. Lett.} {\bf 2011}, {\em 106},~ doi:10.1103/PhysRevLett.106.186406.

\bibitem[Del Campo \em{et~al.}(2012)del Campo, Gazquez, Trickey, and
 Vela]{pbemol}
Del Campo, J.M.; Gazquez, J.L.; Trickey, S.B.; Vela, A.
\newblock Non-empirical improvement of PBE and its hybrid PBE0 for general
 description of molecular properties.
\newblock {\em J. Chem. Phys.} {\bf 2012}, {\em 136}, doi:10.1063/1.3691197.

\bibitem[Peverati and Truhlar(2012)]{peverati12}
Peverati, R.; Truhlar, D.G.
\newblock Exchange-Correlation Functional with Good Accuracy for Both
 Structural and Energetic Properties while Depending Only on the Density and
 Its Gradient.
\newblock {\em J. Chem. Theory Comput.} {\bf 2012}, {\em 8},~2310--2319.

\bibitem[Armiento and Mattsson(2005)]{am05}
Armiento, R.; Mattsson, A.E.
\newblock Functional designed to include surface effects in self-consistent
 density functional theory.
\newblock {\em Phys. Rev. B} {\bf 2005}, {\em 72},~ doi:10.1103/PhysRevB.72.085108.

\bibitem[Perdew \em{et~al.}(2008)Perdew, Ruzsinszky, Csonka, Vydrov, Scuseria,
 Constantin, Zhou, and Burke]{pbesol}
Perdew, J.P.; Ruzsinszky, A.; Csonka, G.I.; Vydrov, O.A.; Scuseria, G.E.;
 Constantin, L.A.; Zhou, X.; Burke,~K.
\newblock Restoring the Density-Gradient Expansion for Exchange in Solids and
 Surfaces.
\newblock {\em Phys. Rev. Lett.} {\bf 2008}, {\em 100}, doi:10.1103/PhysRevLett.100.136406.

\bibitem[Vela \em{et~al.}(2012)Vela, Pacheco-Kato, Gázquez, del Campo, and
 Trickey]{vmt}
Vela, A.; Pacheco-Kato, J.C.; Gázquez, J.L.; del Campo, J.M.; Trickey, S.B.
\newblock Improved constraint satisfaction in a simple generalized gradient
 approximation exchange functional.
\newblock {\em J. Chem. Phys.} {\bf 2012}, {\em 136}, doi:10.1063/1.3701132.

\bibitem[Fabiano \em{et~al.}(2010)Fabiano, Constantin, and {Della
 Sala}]{pbeint}
Fabiano, E.; Constantin, L.A.; {Della Sala}, F.
\newblock Generalized gradient approximation bridging the rapidly and slowly
 varying density regimes: A PBE-like functional for hybrid interfaces.
\newblock {\em Phys. Rev. B} {\bf 2010}, {\em 82},~doi:10.1103/PhysRevB.82.113104.

\bibitem[Constantin \em{et~al.}(2012)Constantin, Fabiano, and {Della
 Sala}]{zpbeint}
Constantin, L.A.; Fabiano, E.; {Della Sala}, F.
\newblock Spin-dependent gradient correction for more accurate atomization
 energies of molecules.
\newblock {\em J. Chem. Phys.} {\bf 2012}, {\em 137},~doi:10.1063/1.4766324.

\bibitem[Chiodo \em{et~al.}(2012)Chiodo, Constantin, Fabiano, and {Della
 Sala}]{q2D}
Chiodo, L.; Constantin, L.A.; Fabiano, E.; {Della Sala}, F.
\newblock Nonuniform scaling applied to surface energies of transition metals.
\newblock {\em Phys. Rev. Lett.} {\bf 2012}, {\em 108},~doi:10.1103/PhysRevLett.108.126402.

\bibitem[Peverati \em{et~al.}(2011)Peverati, Zhao, and Truhlar]{sogga11}
Peverati, R.; Zhao, Y.; Truhlar, D.G.
\newblock Generalized gradient approximation that recovers the second-order
 density-gradient expansion with optimized across-the-board performance.
\newblock {\em J. Phys. Chem. Lett.} {\bf 2011}, {\em 2},~1991--1997.

\bibitem[Constantin \em{et~al.}(2016)Constantin, Terentjevs, {Della Sala},
 Cortona, and Fabiano]{sg4}
Constantin, L.A.; Terentjevs, A.; {Della Sala}, F.; Cortona, P.; Fabiano, E.
\newblock Semiclassical atom theory applied to solid-state physics.
\newblock {\em Phys. Rev. B} {\bf 2016}, {\em 93},~doi:10.1103/PhysRevB.93.04512626.

\bibitem[Tao \em{et~al.}(2003)Tao, Perdew, Staroverov, and Scuseria]{tpss}
Tao, J.; Perdew, J.P.; Staroverov, V.N.; Scuseria, G.E.
\newblock Climbing the density functional ladder: Nonempirical
 Meta\char21{}Generalized gradient approximation designed for Mo lecules and
 solids.
\newblock {\em Phys. Rev. Lett.} {\bf 2003}, {\em 91}, doi:10.1103/PhysRevLett.91.146401.

\bibitem[Perdew \em{et~al.}(2009)Perdew, Ruzsinszky, Csonka, Constantin, and
 Sun]{revtpss}
Perdew, J.P.; Ruzsinszky, A.; Csonka, G.I.; Constantin, L.A.; Sun, J.
\newblock Workhorse semilocal density functional for condensed matter physics
 and quantum chemistry.
\newblock {\em Phys. Rev. Lett.} {\bf 2009}, {\em 103}, doi:10.1103/Phys RevLett.103.026403.

\bibitem[Sun \em{et~al.}(2012)Sun, Xiao, and Ruzsinszky]{mggams1}
Sun, J.; Xiao, B.; Ruzsinszky, A.
\newblock Communication: Effect of the orbital-overlap dependence in the meta
 generalized gradient approximation.
\newblock {\em J. Chem. Phys.} {\bf 2012}, {\em 137},~doi:10.1063/1.4742312.

\bibitem[Sun \em{et~al.}(2015)Sun, Perdew, and Ruzsinszky]{sunpnas}
Sun, J.; Perdew, J.P.; Ruzsinszky, A.
\newblock Semilocal density functional obeying a strongly tightened bound for~exchange.
\newblock {\em Proc. Nat. Acad. Sci. USA} {\bf 2015}, {\em 112},~685--689.

\bibitem[Zhao and Truhlar(2006)]{m06l}
Zhao, Y.; Truhlar, D.G.
\newblock A new local density functional for main-group thermochemistry,
 transition metal bonding, thermochemical kinetics, and noncovalent
 interactions.
\newblock {\em J. Chem. Phys.} {\bf 2006}, {\em 125}, doi:10.1063/1.2370993.

\bibitem[Peverati and Truhlar(2012)]{m11l}
Peverati, R.; Truhlar, D.G.
\newblock M11-L: A local density functional that provides improved accuracy for
 electronic structure calculations in chemistry and physics.
\newblock {\em J. Phys. Chem. Lett.} {\bf 2012}, {\em 3},~117--124.

\bibitem[Constantin \em{et~al.}(2013)Constantin, Fabiano, and {Della
 Sala}]{bloc}
Constantin, L.A.; Fabiano, E.; {Della Sala}, F.
\newblock Meta-GGA exchange-correlation functional with a balanced treatment of
 nonlocality.
\newblock {\em J. Chem. Theory Comput.} {\bf 2013}, {\em 9},~2256--2263.

\bibitem[Mardirossian and Head-Gordon(2015)]{b97mv2015}
Mardirossian, N.; Head-Gordon, M.
\newblock Mapping the genome of meta-generalized gradient approximation density
 functionals: The search for B97M-V.
\newblock {\em J. Chem. Phys.} {\bf 2015}, {\em 142},~doi:10.1063/1.4907719.

\bibitem[del Campo \em{et~al.}(2012)del Campo, Gazquez, Trickey, and
 Vela]{vt84}
Del Campo, J.M.; Gazquez, J.L.; Trickey, S.; Vela, A.
\newblock A new meta-GGA exchange functional based on an~improved
 constraint-based GGA.
\newblock {\em Chem. Phys. Lett.} {\bf 2012}, {\em 543},~179--183.

\bibitem[Peverati and Truhlar(2012)]{mn12l}
Peverati, R.; Truhlar, D.G.
\newblock An improved and broadly accurate local approximation to the
 exchange-correlation density functional: The MN12-L functional for electronic
 structure calculations in chemistry and physics.
\newblock {\em Phys. Chem. Chem. Phys.} {\bf 2012}, {\em 14},~13171--13174.

\bibitem[Wellendorff \em{et~al.}(2014)Wellendorff, Lundgaard, Jacobsen, and
 Bligaard]{beefmeta14}
Wellendorff, J.; Lundgaard, K.T.; Jacobsen, K.W.; Bligaard, T.
\newblock mBEEF: An accurate semi-local Bayesian error estimation density
 functional.
\newblock {\em J. Chem. Phys.} {\bf 2014}, {\em 140},~doi:10.1063/1.4870397.

\bibitem[Sun \em{et~al.}(2015)Sun, Ruzsinszky, and Perdew]{scan}
Sun, J.; Ruzsinszky, A.; Perdew, J.P.
\newblock Strongly constrained and appropriately normed semilocal density~functional.
\newblock {\em Phys. Rev. Lett.} {\bf 2015}, {\em 115},~doi:10.1103/PhysRevLett.115.036402.

\bibitem[Becke(1983)]{Beckeh}
Becke, A.D.
\newblock Hartree-Fock exchange energy of an inhomogeneous electron gas.
\newblock {\em Int. J. Quantum Chem.} {\bf 1983}, {\em 23},~1915--1922.

\bibitem[Armiento and K\"ummel(2013)]{arm13}
Armiento, R.; K\"ummel, S.
\newblock Orbital localization, charge transfer, and band gaps in semilocal
 density- functional~theory.
\newblock {\em Phys. Rev. Lett.} {\bf 2013}, {\em 111},~doi:10.1103/PhysRevLett.111.036402.

\bibitem[Constantin \em{et~al.}(2013)Constantin, Fabiano, and {Della
 Sala}]{blochole}
Constantin, L.A.; Fabiano, E.; {Della Sala}, F.
\newblock Construction of a general semilocal exchange-correlation hole~model:
 Application to nonempirical meta-GGA functionals.
\newblock {\em Phys. Rev. B} {\bf 2013}, {\em 88}, doi:10.1103/Phys RevB.88.125112.

\bibitem[Perdew \em{et~al.}(1996)Perdew, Burke, and Wang]{perdewPRB96}
Perdew, J.P.; Burke, K.; Wang, Y.
\newblock Generalized gradient approximation for the exchange-correlation hole
 of \linebreak a many-electron system.
\newblock {\em Phys. Rev. B} {\bf 1996}, {\em 54},~doi:10.1103/PhysRevB.54.16533.

\bibitem[Ernzerhof and Perdew(1998)]{ernzerhofJCP98}
Ernzerhof, M.; Perdew, J.P.
\newblock Generalized gradient approximation to the angle-and system-averaged
 exchange~hole.
\newblock {\em J. Chem. Phys.} {\bf 1998}, {\em 109},~3313--3320.

\bibitem[Vydrov \em{et~al.}(2006)Vydrov, Heyd, Krukau, and
 Scuseria]{vydrov2006importance}
Vydrov, O.A.; Heyd, J.; Krukau, A.V.; Scuseria, G.E.
\newblock Importance of short-range versus long-range Hartree-Fock exchange for
 the performance of hybrid density functionals.
\newblock {\em J. Chem. Phys.} {\bf 2006}, {\em 125}, doi:10.1063/1.2244560.

\bibitem[Tao(2001)]{tao01}
Tao, J.
\newblock Exchange energy density of an atom as a functional of the electron
 density.
\newblock {\em J. Chem. Phys.} {\bf 2001}, {\em 115},~3519--3530.

\bibitem[Cancio \em{et~al.}(2012)Cancio, Wagner, and Wood]{cancioIJQC12}
Cancio, A.C.; Wagner, C.E.; Wood, S.A.
\newblock Laplacian-based models for the exchange energy.
\newblock {\em Int. J. Quantum Chem.} {\bf 2012}, {\em 112},~3796--3806.

\bibitem[Lembarki and Chermette(1994)]{lc94}
Lembarki, A.; Chermette, H.
\newblock Obtaining a gradient-corrected kinetic-energy functional from the
 Perdew-Wang exchange functional.
\newblock {\em Phys. Rev. A} {\bf 1994}, {\em 50},~5328--5331.

\bibitem[Tran and Wesolowski(2002)]{tw02}
Tran, F.; Wesoloski, T.A.
\newblock Link between the kinetic- and exchange-energy functionals in the
 generalized gradient approximation.
\newblock {\em Int. J. Quantum Chem.} {\bf 2002}, {\em 89},~doi:10.1002/qua.10306.

\bibitem[Lee \em{et~al.}(1991)Lee, Lee, and Parr]{llp91}
Lee, H.; Lee, C.; Parr, R.G.
\newblock Conjoint gradient correction to the Hartree-Fock kinetic- and
 exchange-energy density functionals.
\newblock {\em Phys. Rev. A} {\bf 1991}, {\em 44},~768--771.

\bibitem[Thakkar(1992)]{thak92}
Thakkar, A.J.
\newblock Comparison of kinetic-energy density functionals.
\newblock {\em Phys. Rev. A} {\bf 1992}, {\em 46},~6920--6924.

\bibitem[Laricchia \em{et~al.}(2011)Laricchia, Fabiano, Constantin, and {Della
 Sala}]{apbek}
Laricchia, S.; Fabiano, E.; Constantin, L.A.; {Della Sala}, F.
\newblock Generalized gradient approximations of the noninteracting kinetic
 energy from the semiclassical atom theory: Rationalization of the accuracy of
 the frozen density embedding theory for nonbonded interactions.
\newblock {\em J. Chem. Theory Comput.} {\bf 2011}, {\em 7},~2439--2451.

\bibitem[Laricchia \em{et~al.}(2014)Laricchia, Constantin, Fabiano, and {Della
 Sala}]{fde_lap}
Laricchia, S.; Constantin, L.A.; Fabiano, E.; {Della Sala}, F.
\newblock Laplacian-Level kinetic energy approximations based on the
 fourth-order gradient expansion: Global assessment and application to the
 subsystem formulation of density functional theory.
\newblock {\em J. Chem. Theory Comput.} {\bf 2014}, {\em 10},~164--179.

\bibitem[Tran and Wesolowski(2013)]{weso_chap_funct}
Tran, F.; Wesolowski, T.A.
\newblock Semilocal Approximation for the Kinetic Energy. In {\em Recent
 Advances in Computational Chemistry 6}; Wesolowski, T.A., Wang, Y.A., Eds.;
 World Scientific: Singapore, 2013; \linebreak pp. 429--442.

\bibitem[Garc\'{\i}a-Aldea and Alvarellos(2008)]{alva08}
Garc\'{\i}a-Aldea, D.; Alvarellos, J.E.
\newblock Approach to kinetic energy density functionals: Nonlocal terms with
 the structure of the von Weizs\"acker functional.
\newblock {\em Phys. Rev. A} {\bf 2008}, {\em 77},~doi:10.1103/PhysRevA.77.022502.

\bibitem[Karasiev \em{et~al.}(2009)Karasiev, Jones, Trickey, and
 Harris]{kara09}
Karasiev, V.V.; Jones, R.S.; Trickey, S.B.; Harris, F.E.
\newblock Properties of constraint-based single-point approximate kinetic
 energy functionals.
\newblock {\em Phys. Rev. B} {\bf 2009}, {\em 80},~doi:10.1103/PhysRevB.80.245120120.

\bibitem[{Garcia Lastra} \em{et~al.}(2008){Garcia Lastra}, Kaminski, and
 Wesolowski]{lastra08}
{Garcia Lastra}, J.M.; Kaminski, J.W.; Wesolowski, T.A.
\newblock Orbital-free effective embedding potential at nuclear~cusps.
\newblock {\em J. Chem. Phys.} {\bf 2008}, {\em 129},~074107.

\bibitem[Perdew and Constantin(2007)]{MGGA}
Perdew, J.P.; Constantin, L.A.
\newblock Laplacian-level density functionals for the kinetic energy density
 and exchange-correlation energy.
\newblock {\em Phys. Rev. B} {\bf 2007}, {\em 75},~doi:10.1103/PhysRevB.75.155109.

\bibitem[Karasiev \em{et~al.}(2013)Karasiev, Chakraborty, Shukruto, and
 Trickey]{kara13}
Karasiev, V.V.; Chakraborty, D.; Shukruto, O.A.; Trickey, S.B.
\newblock Nonempirical generalized gradient approximation free-energy
 functional for orbital-free simulations.
\newblock {\em Phys. Rev. B} {\bf 2013}, {\em 88},~doi:10.1103/Phys RevB.88.161108.

\bibitem[{Della Sala} \em{et~al.}(2015){Della Sala}, Fabiano, and
 Constantin]{alpha}
{Della Sala}, F.; Fabiano, E.; Constantin, L.A.
\newblock Kohn--Sham kinetic energy density in the nuclear and asymptotic
 regions: Deviations from the von Weizs\"acker behavior and applications to
 density functionals.
\newblock {\em Phys. Rev. B} {\bf 2015}, {\em 91},~doi:10.1103/PhysRevB.91.035126.

\bibitem[Qian(2007)]{qian07}
Qian, Z.
\newblock Exchange and correlation near the nucleus in density functional
 theory.
\newblock {\em Phys. Rev. B} {\bf 2007}, {\em 75}, doi:10.1103/PhysRevB.75.193104.

\bibitem[Nagy and March(1989)]{nagy89}
Nagy, A.; March, N.H.
\newblock Exact potential-phase relation for the ground state of the C atom.
\newblock {\em Phys. Rev. A} {\bf 1989}, {\em 40},~554--557.

\bibitem[Santamaria and March(1990)]{santamaria90}
Santamaria, R.; March, N.
\newblock Kinetic energy density as a function of subshell electron densities.
\newblock {\em J. Mol. Struct. } {\bf 1990}, {\em 205},~35--41.

\bibitem[Zhou and Chu(2005)]{zhou05}
Zhou, Z.; Chu, S.I.
\newblock Spin-dependent localized Hartree-Fock density-functional calculation
 of singly, doubly, and triply excited and Rydberg states of He- and Li-like
 ions.
\newblock {\em Phys. Rev. A} {\bf 2005}, {\em 71}, doi:10.1103/Phys RevA.71.022513.

\bibitem[Heilmann and Lieb(1995)]{Lieb}
Heilmann, O.J.; Lieb, E.H.
\newblock Electron density near the nucleus of a large atom.
\newblock {\em Phys. Rev. A} {\bf 1995}, {\em 52},~3628--3643.

\bibitem[Constantin \em{et~al.}(2010)Constantin, Snyder, Perdew, and
 Burke]{ioni}
Constantin, L.A.; Snyder, J.C.; Perdew, J.P.; Burke, K.
\newblock Communication: Ionization potentials in the limit of large atomic
 number.
\newblock {\em J. Chem. Phys.} {\bf 2010}, {\em 133},~doi:10.1063/1.3522767.

\bibitem[Howard \em{et~al.}(2001)Howard, March, and Van~Doren]{march01}
Howard, I.A.; March, N.H.; Van~Doren, V.E.
\newblock \textit{r}- and \textit{p}-space electron densities and related
 kinetic and exchange energies in terms of \textit{s} states alone for the
 leading term in the $1/Z$ expansion for nonrelativistic closed-shell atomic
 ions.
\newblock {\em Phys. Rev. A} {\bf 2001}, {\em 63},~doi:10.1103/PhysRevA.63.0625011.

\bibitem[March and Nagy(2010)]{march10}
March, N.H.; Nagy, A.
\newblock Pauli potential in terms of kinetic energy density and electron
 density in the leading Coulombic term of the nonrelativistic $1/Z$ expansion
 of spherical atomic ions.
\newblock {\em Phys. Rev. A} {\bf 2010}, {\em 81}, doi:10.1103/PhysRevA.81.014502.

\bibitem[Bog\'ar \em{et~al.}(2011)Bog\'ar, Bartha, Bartha, and March]{march11}
Bog\'ar, F.; Bartha, F.; Bartha, F.A.; March, N.H.
\newblock Pauli potential from Heilmann-Lieb electron density obtained by
 summing hydrogenic closed-shell densities over the entire bound-state
 spectrum.
\newblock {\em Phys. Rev. A} {\bf 2011}, {\em 83}, doi:10.1103/PhysRevA.83.014502.

\bibitem[Lee \em{et~al.}(2009)Lee, Constantin, Perdew, and Burke]{LCPB09}
Lee, D.; Constantin, L.A.; Perdew, J.P.; Burke, K.
\newblock Condition on the Kohn--Sham kinetic energy and modern parametrization
 of the Thomas--Fermi density.
\newblock {\em J. Chem. Phys.} {\bf 2009}, {\em 130},~doi:10.1063/1.3059783.

\bibitem[Elliott \em{et~al.}(2008)Elliott, Lee, Cangi, and Burke]{burkesemi}
Elliott, P.; Lee, D.; Cangi, A.; Burke, K.
\newblock Semiclassical Origins of Density Functionals.
\newblock {\em Phys. Rev. Lett.} {\bf 2008}, {\em 100}, doi:10.1103/PhysRevLett.100.256406.

\bibitem[Kato(1957)]{kato}
Kato, T.
\newblock On the eigenfunctions of many-particle systems in quantum mechanics.
\newblock {\em Commun. Pure Appl. Math.} {\bf 1957}, {\em
 10},~151--177.

\bibitem[Liu \em{et~al.}(1995)Liu, Parr, and Nagy]{liu95}
Liu, S.; Parr, R.G.; Nagy, A.
\newblock Cusp relations for local strongly decaying properties in electronic
 systems.
\newblock \linebreak {\em Phys. Rev. A} {\bf 1995}, {\em 52},~2645--2651.

\bibitem[Horowitz \em{et~al.}(2009)Horowitz, Constantin, Proetto, and
 Pitarke]{CP3}
Horowitz, C.M.; Constantin, L.A.; Proetto, C.R.; Pitarke, J.M.
\newblock Position-dependent exact-exchange energy for slabs and semi-infinite
 jellium.
\newblock {\em Phys. Rev. B} {\bf 2009}, {\em 80},~235101.

\bibitem[Engel, E and Vosko, SH]{engel1}
Engel, E.; Vosko, S.H.
\newblock Accurate optimized-potential-model solutions for spherical spin-polarized atoms: Evidence for limitations of the
exchange-only local spin-density and generalized-gradient approximations
\newblock {\em Phys. Rev. A} {\bf 1993}, {\em 47}, doi:10.1103/PhysRevA.47.2800.

\bibitem[Engel, E]{engel2}
Engel, E.
\newblock Orbital-dependent functionals for the exchange-correlation energy: A third generation of density functionals.\newblock In {\em A Primer in Density Functional Theory}; Springer: Berlin/Heidelberg, Germany, 2003; pp. 56--122.

\bibitem[March(1986)]{spatgen}
March, N.H.
\newblock Spatially dependent generalization of Kato's theorem for atomic
 closed shells in a bare Coulomb~field.
\newblock {\em Phys. Rev. A} {\bf 1986}, {\em 33},~88--89.

\bibitem[Seidl \em{et~al.}(1996)Seidl, G\"orling, Vogl, Majewski, and
 Levy]{seidl96}
Seidl, A.; G\"orling, A.; Vogl, P.; Majewski, J.A.; Levy, M.
\newblock Generalized Kohn--Sham schemes and the band-gap~problem.
\newblock {\em Phys. Rev. B} {\bf 1996}, {\em 53},~3764--3774.

\bibitem[Talman and Shadwick(1976)]{talman76}
Talman, J.D.; Shadwick, W.F.
\newblock Optimized effective atomic central potential.
\newblock {\em Phys. Rev. A} {\bf 1976}, {\em 14}, doi:10.1103/ PhysRevA.14.36.

\bibitem[K\"ummel and Kronik(2008)]{kummelrev}
K\"ummel, S.; Kronik, L.
\newblock Orbital-dependent density functionals: Theory and applications.
\newblock {\em Rev. Mod. Phys.} {\bf 2008}, {\em 80},~3--60.

\bibitem[{Della Sala} and G\"orling(2001)]{lhf}
{Della Sala}, F.; G\"orling, A.
\newblock Efficient localized Hartree-Fock methods as effective exact-exchange
 Kohn–Sham methods for molecules.
\newblock {\em J. Chem. Phys.} {\bf 2001}, {\em 115}, doi:10.1063/1.13980938.

\bibitem[Perdew \em{et~al.}(2008)Perdew, Staroverov, Tao, and Scuseria]{hyper1}
Perdew, J.P.; Staroverov, V.N.; Tao, J.; Scuseria, G.E.
\newblock Density functional with full exact exchange, balanced nonlocality of
 correlation, and constraint satisfaction.
\newblock {\em Phys. Rev. A} {\bf 2008}, {\em 78}, doi:10.1103/Phys RevA.78.052513.

\bibitem[Odashima and Capelle(2009)]{hyper2}
Odashima, M.M.; Capelle, K.
\newblock Nonempirical hyper-generalized-gradient functionals constructed from
 the Lieb-Oxford bound.
\newblock {\em Phys. Rev. A} {\bf 2009}, {\em 79}, doi:10.1103/PhysRevA.79.062515 .

\bibitem[Haunschild \em{et~al.}(2012)Haunschild, Odashima, Scuseria, Perdew,
 and Capelle]{hyper3}
Haunschild, R.; Odashima, M.M.; Scuseria, G.E.; Perdew, J.P.; Capelle, K.
\newblock Hyper-generalized-gradient functionals constructed from the
 Lieb-Oxford bound: Implementation via local hybrids and thermochemical
 assessment.
\newblock {\em J. Chem. Phys.} {\bf 2012}, {\em 136}, doi:10.1063/1.4712017.
\end{thebibliography}
\end{document}